\documentclass[aps,prb,preprint,groupedaddress,showpacs]{revtex4}

\usepackage{graphics}

\begin{document}



\title{
Quantum-fluctuation-induced collisions
and subsequent excitation gap
of an elastic string between walls
}


\author{Yoshihiro Nishiyama}
\email[]{nisiyama@psun.phys.okayama-u.ac.jp}
\affiliation{Department of Physics, Faculty of Science,
Okayama University, Okayama 700-8530, Japan.}


\date{\today}

\begin{abstract}
An elastic string embedded between rigid walls
is simulated by means of the density-matrix
renormalization group.
The string collides against the walls
owing to the quantum-mechanical zero-point fluctuations.
Such ``quantum entropic'' interaction has come under thorough
theoretical investigation
in the context of
the stripe phase observed experimentally in doped cuprates.
We found that the excitation gap opens 
in the form of exponential singularity
$\Delta E \sim \exp(-Ad^\sigma)$ ($d$: wall spacing)
with the exponent $\sigma =0.6(3)$,
which is substantially smaller than the meanfield value $\sigma=2$.
That is, the excitation gap is much larger than
that anticipated from meanfield, suggesting that the string
is subjected to robust pinning potential due to the
quantum collisions.
This feature supports Zaanen's
``order out of disorder'' mechanism which would be responsible
to the stabilization of the stripe phase.
\end{abstract}


\maketitle

\section{\label{section_1}Introduction}

Recently, Zaanen brought up a problem of ``quantum string,''
which is a linelike object subjected to line tension,
and it wanders owing to quantum zero-point fluctuations 
\cite{Zaanen00,Mukhin01}.
Central concern is to estimate the interaction among strings
as it wanders quantum-mechanically and undergoes entropy-reducing
collisions with adjacent neighbors.
Statistical mechanics of quantum-string gas would be responsible to
the low-energy physics of the stripe phase observed experimentally in
doped cuprates \cite{Tranquada95,Wakimoto99,Noda99}.
In particular,
one is motivated to gain insights how the stripe pattern
acquires stability.
Actually, a good deal of theoretical analyses
had predicted tendency toward stripe-pattern formation
\cite{Zaanen89,Poilblanc89,Schulz90,Kato90}.
However,
{\it ab initio} simulations on the $t$-$J$ model
still remain controversial about that issue
\cite{Prelovsek93,White98,Hellberg99,White00,Hellberg00}.

In the series of papers \cite{Zaanen00,Mukhin01}, the authors pointed out that
the ``quantum entropic'' interaction lies out of
conventional picture, and it would rather give rise
to the solidification of the gas of strings;
namely, the stripe phase is stabilized by mutual collisions.
(In a different context, the entropic interaction was explored 
in Refs. \onlinecite{Bricmont86,Sornette86}.)
Their analysis \cite{Zaanen00,Mukhin01} is based on 
the Helfrich approximation \cite{Helfrich84},
which has been very successful
in the course of studies of stacked membranes under
{\em thermal} undulations.
Based on this approximation, they reveled
a significance of 
{\em long}-wavelength fluctuations.
Note that in the conventional picture, on the contrary,
the string is ``disordered'' as in the Einstein-like
view of crystal, and
only short-wavelength fluctuations contribute to collisions.

To be specific,
the key relation in their theory is the following,
\begin{equation}
\label{SC_condition}
f = C \frac{B}{\Sigma d^2} \left(
          \log \left( \frac{\Sigma d}{B} \right)
               + C' \right) ,
\end{equation}
with collision-induced energy cost $f$ (per unit volume)
and elasticity modulus (with respect to the compression of string intervals)
$B=d^2 \partial^2 f/\partial d^2$; see the Hamiltonian (\ref{Hamiltonian})
as well.
The logarithmic term signals the significance of 
long-wavelength fluctuations.
The relation (\ref{SC_condition}) yields remarkable consequences.
For instance,
the elastic modulus is given by the {\em stretched} exponential form
$B\sim \exp(-Ad^{2/3})$ \cite{Zaanen00,Mukhin01}.
It is noteworthy that the elasticity modulus
$B$ is much larger than that of
meanfield $B\sim\exp(-Ad^2)$.
Hence, the result indicates that contrary to our naive expectations,
short-wavelength modes are suppressed owing to
collisions, and the collisions rather contribute
to the solidification of the string gas;
namely, the stripe phase is stabilized by 
the ``order out of disorder'' mechanism \cite{Zaanen00,Mukhin01}.

In our preceding paper \cite{Nishiyama01}, we have verified the 
above relation (\ref{SC_condition}), and demonstrated that
the elastic modulus is actually governed by the stretched exponential
form $B\sim\exp(-Ad^{0.808(1)})$.
We have carried out the first-principle simulation
by means of the density-matrix renormalization group 
\cite{White92,White93,Peschel99}:
We put a quantum string between
rigid walls with spacing $d$, and measured its repelling interaction.
(This technique has been utilized in the course of studies of 
fluctuation pressure of
stacked membranes
\cite{Janke86,Janke87,Leibler89}.)
The Hamiltonian is given by,
\begin{equation}
\label{Hamiltonian}
{\cal H}=\sum_{i=1}^{L} \left( \frac{p_i^2}{2m}
                                 + V(x_i) \right)
          +\sum_{i=1}^{L-1} \frac{\Sigma}{2} (x_i-x_{i+1})^2    .
\end{equation}
Here, $x_i$ denotes the operator of transverse displacement of
a particle at $i$th site, and $p_i$ is its conjugate momentum.
They satisfy the canonical commutation relations
$[x_i,p_j]={\rm i}\hbar \delta_{ij}$,
$[x_i,x_j]=0$ and
$[p_i,p_j]=0$.
$V(x)$ is the rigid-wall potential with spacing $d$;
\begin{equation}
V(x)=
\left\{
\begin{array}{ll}
  0    &      {\rm for}   \    0 \le x \le d \\
\infty &      {\rm otherwise}
\end{array}
\right. .
\end{equation}
$\Sigma$ denotes the line tension which puts particles into line.
Classical version of this Hamiltonian has been used as a model
for line dislocations and steps on (vicinal) surfaces
\cite{Pokrovsky79,Coppersmith82,Zaanen96}.
Note that for sufficiently large $\Sigma$,
one can take continuum limit, with which one arrives at field-theoretical
version of quantum string \cite{Momoi01}.

In the present paper, we are concerned in the excitation (mass) gap
due to the collisions.
We postulate the following exponential singularity,
\begin{equation}
\label{exponential}
\Delta E \sim \exp(-A d^\sigma) ,
\end{equation}
for sufficiently large $d$.
In Ref. \onlinecite{Mukhin01}, the gap (crossover temperature)
$T_0$ is calculated in the form
$T_0 \propto \sqrt{B}$; namely, one should obtain
$\sigma=2/3$.
This result cannot be understood in terms of the meanfield picture
yielding much smaller mass gap 
$\sigma=2$.
(We will outline this picture in the next section.)
The purpose of this paper is to judge the validity of those
scenarios by performing first-principle simulations.
Our result is $\sigma \approx 0.6(3)$ for $\Sigma=0.5$-$4$.

The rest of this paper is organized as follows.
In the next section, we argue physical implications of
the singularity exponent $\sigma$.
We introduce the viewpoint of critical phenomena in order to
interpret the gap formula (\ref{exponential}).
The meanfield argument is also explicated.
In Section \ref{section_3}, we perform numerical simulations,
and
show evidences of the breakdown of meanfield.
In the last section, we give summary and discussions.

\section{
\label{section_2}
Interpretation of the singularity exponent $\sigma$
---
a statistical-mechanical overview
}

In Introduction, we have overviewed the quantum string 
in the context of the stripe phase.
In this section, we will survey different aspects of the quantum string
in terms of statistical mechanics.
This viewpoint is utilized in the simulation-data analyses 
in the succeeding section \ref{section_3}.

In the path-integral picture, a quantum string spans a 
``world sheet'' as time evolves \cite{Fradkin83}.
Therefore, a quantum string is equivalent to the random surface
under {\em thermal} undulations.
The random surface is in the critical phase of Kosterlitz and Thouless.
(In our model (\ref{Hamiltonian}), however, there appears no ``flat phase,''
because the ``height'' variables $\{ x_i \}$ are continuous.)
The presence of rigid walls is supposed to destroy the KT criticality.
Therefore, the collision-induced mass gap (\ref{exponential}) reflects
the universality class of the phase transition at $1/d\to0$.
In that sense, our aim is to determine the universality class.

The phase transition has not been studied very extensively so far.
The rigid-wall potential is, by nature, non-analytic,
and hence, it is quite cumbersome to carry out perturbative analyses
such as loop expansions.
Moreover, the confinement due to walls destroys the applicability 
of the duality transformation, with which one could establish
 an elucidating equivalence
between the random surface and the two-dimensional $XY$ model.

Even in computer simulation, the verification of the gap formula
(\ref{exponential}) is rather troublesome.
Note that the gap opens very slowly.
That is, 
the correlation length is kept exceedingly large in the vicinity of
the critical point.
The correlation length would exceed the system sizes.
Note that
the conventional criticality exhibits the {\em power law} singularity,
\begin{equation}
\label{power}
\Delta E \sim \xi^{-1} \sim (1/d)^{\nu} ,
\end{equation}
which would be much easier to cope with.
We will overcome the difficulty through resorting to the
density-matrix renormalization group \cite{White92,White93,Peschel99};
with this technique, we treat system sizes up to $L=42$.

Now,
we are in a position to discuss the meaning of
the singularity exponent $\sigma$.
The correlation length is proportional to the inverse of
the excitation (mass) gap; namely, $\xi \sim \exp(Ad^\sigma)$.
$\xi$ sets the characteristic 
length scale of the collision intervals.
Therefore, in the length scale $l=\xi$,
the random surface fluctuates freely.
And so,
the mean fluctuation deviation is calculated,
\begin{equation} 
\label{deviation_and_exponent}
\sqrt{\langle x_i^2 \rangle} \sim (\log l)^{1/2} \sim d^{\sigma/2}.
\end{equation}
From this relation, we see that $\sigma$ reflects an amount of
fluctuation amplitudes.
In the meantime, the meanfield picture insists 
$\sqrt{\langle x_i^2 \rangle} \propto d$.
(More precisely, an artificial mass term is included so as to
enforce the condition; see Refs. \onlinecite{Janke86,Sornette86}.)
Hence, one should obtain,
\begin{equation}
\sigma=2 ,
\end{equation}
in the meanfield picture.

On the contrary, as is mentioned in Introduction,
recent theory \cite{Zaanen00,Mukhin01} 
predicts much smaller
singularity exponent $\sigma=2/3$,
from which one obtains a counterintuitive result
$\sqrt{\langle x_i^2\rangle} \sim d^{1/3} \ll d$.
The result indicates that the string is straightened 
macroscopically by collisions:
This is quite reminiscent of the aforementioned claim 
that the infrared modes are still active
withstanding the collisions \cite{Zaanen00,Mukhin01}.
Similar, but slightly modest, exponent
$\sigma=1$ was reported
(in a different context) with analytical calculations
\cite{Bricmont86,Sornette86}.

\section{\label{section_3}Numerical results and discussions}

In this section, we perform first-principle simulations so as to
estimate the singularity exponent $\sigma$.
First, we will explicate the simulation algorithm.

\subsection{Details of 
the density-matrix renormalization group}

The density-matrix renormalization group has been utilized
for the purpose of treating very large system sizes \cite{White92,White93}.
The technique is based on an elaborate reduction of
the Hilbert-space bases, and it has been applied to various
spin and electron systems very successfully \cite{Peschel99}.
Meanwhile, the technique had become applicable to bosonic systems
such as phonons and oscillators
\cite{Caron96,Zhang98,Nishiyama99,Nishiyama02}.
(Note that the full exact diagonalization does not apply,
because even a single oscillator spans infinite-dimensional Fock space.
The huge dimensionality overwhelms the computer memory space.)
Here, we employ the density-matrix-renormalization
technique in order to diagonalize the quantum string
(\ref{Hamiltonian}).
Full account of technical details is presented in our preceding paper
\cite{Nishiyama01}.
In this paper,
we will outline the simulation algorithm with an emphasis on 
the modification to calculate the excitation gap precisely.

To begin with, we need to set up local (on site) Hilbert-space bases:
Provided that the line tension $\Sigma$ is turned off, 
oscillators are independent,
 and each of them
reduces to the text-book problem
of ``particle in a box.''
Hence, the full set of eigensystems is calculated easily.
In this way,
we set up the intra-site Hilbert space with use of the low-lying $M=12$
states, and discarded the others.

Secondly, provided that those intra-site bases are at hand,
we are able to apply the density-matrix renormalization group.
The total system consists of block, site, site and block.
The left-half part (block and site) is renormalized into a new block
with reduced dimensionality $m$; we set, at most, $m=50$.
More precisely, 
those $m$ bases are chosen from the eigenvectors of the density matrix
for the left-half part,
\begin{equation}
\rho = {\rm Tr}_{\rm R} | \Psi \rangle\langle \Psi |   ,
\end{equation}
with the largest $m$ eigenvalues.
${\rm Tr}_{\rm R}$ denotes the partial trace with respect to
 the right-half part,
and $|\Psi\rangle=|0\rangle+|1\rangle$ 
with the ground state $|0\rangle$ and 
the first excitation state $|1\rangle$ of the total system.
Note that in our previous simulation \cite{Nishiyama01},
we simply set $|\Psi\rangle=|0\rangle$, because we were just concerned in the
ground state properties.

To summarize, we had carried out truncations of bases through two steps:
One is the truncation of the intra-site bases $M$, and the other is the
truncation of the ``block'' states $m$ 
through the density-matrix renormalization.
Each of these procedures is monitored carefully, 
and the relative error of low-lying energy levels is kept within $10^{-7}$;
see Ref. \onlinecite{Nishiyama01} for details.

We have repeated renormalizations 20-times. 
The system size reaches $L=42$.


\subsection{Excitation energy gap}

\begin{figure}
\includegraphics{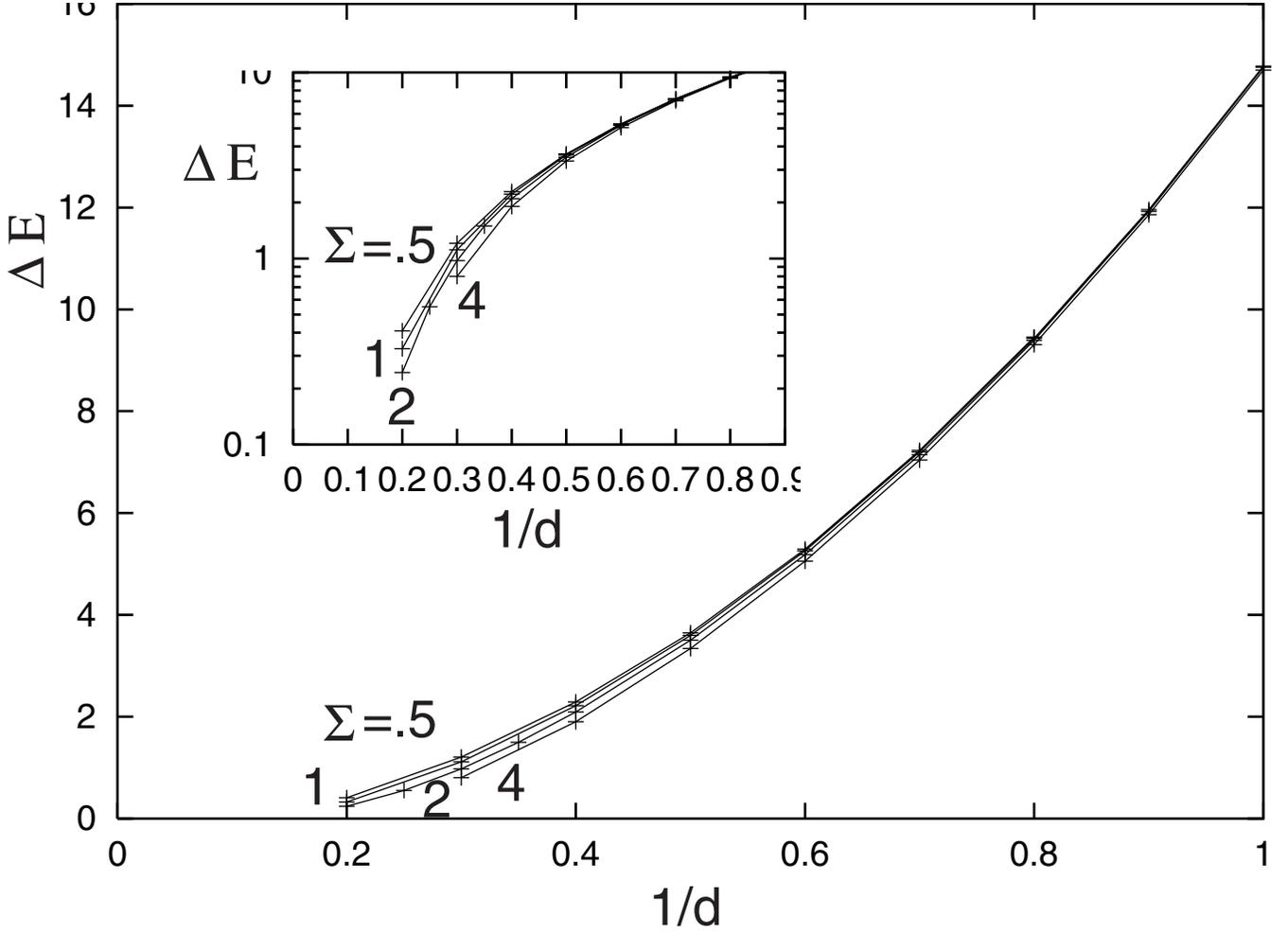}%
\caption{\label{figure_gap}
Excitation gap $\Delta E$ is plotted for $1/d$ 
($d$: wall spacing).
The excitation gap $\Delta E$ opens extremely slowly:
This feature is characteristic of the universality
class of exponential singularity (\ref{exponential})
rather than the power law (\ref{power}).
The inset shows the logarithmic plot. 
The curves bend convexly.
Therefore,
the singularity exponent would not exceed
$\sigma = 1$.
}
\end{figure}

We have plotted the energy gap against the inverse of wall spacing $1/d$
for several values of
line tension
$\Sigma$; see
Fig. \ref{figure_gap}.
From the plot, we see that the gap opens extremely slowly;
namely, in the range $1/d \lesssim 0.3 $, the mass gap $\Delta E$ 
is maintained to be very small.
This is a typical signature of the exponential singularity (\ref{exponential})
rather than the power law (\ref{power}).
For $1/d<0.2$, the simulation does not continue, 
because 
the numerical diagonalization fails in resolving the 
nearly degenerate low-lying levels.

In the inset, we have presented the logarithmic plot.
The data exhibit convex curves.
Hence, we see that the singularity exponent $\sigma$ would
not exceed $\sigma=1$.
As is mentioned in Section \ref{section_2},
the meanfield argument predicts $\sigma=2$.
Therefore, we found that the collision-induced gap cannot be
understood in terms of the
meanfield picture.


\subsection{$\beta$-function analyses}

The criticality is best analyzed 
by the $\beta$-function.
The $\beta$-function describes the flow of 
a certain controllable parameter (in our case, $\delta=1/d$)
with respect to the infinitesimal rescaling of the unit of length;
namely,
\begin{equation}
\beta(1/d)=\frac{\xi(1/d)}{\xi'(1/d)} =
    \left( \frac{\rm d}{{\rm d}(1/d)}\ln \xi \right)^{-1}  ,
\end{equation}
with correlation length $\xi$.
Because the mass gap is the inverse of the correlation length,
we obtain,
\begin{equation}
\label{beta_function}
\label{beta}
        \beta(1/d)    =
 - \left( \frac{\rm d}{{\rm d}(1/d)}\ln \Delta E \right)^{-1},
\end{equation}
with excitation gap $\Delta E$.

As is emphasized in Section \ref{section_2}, 
the gap formula (\ref{exponential}) is
deeply concerned with the criticality at $\delta = 1/d \to 0$.
The $\beta$-function reflects the universality class of 
the phase transition.
For instance,
as for the exponential singularity such as Eq. (\ref{exponential}),
it behaves like,
\begin{equation}
\label{beta_exponential}
\beta(1/d) \sim 1/(A\sigma) (1/d)^{\sigma+1} .
\end{equation} 
On the other hand, the
conventional second-order transition (\ref{power})
is characterized by
the behavior,
\begin{equation}
\label{beta_power}
\beta(1/d) = (1/d)/\nu  .
\end{equation}
In this way, we can read off the 
critical exponents such as $\sigma$ and $\nu$.

As is presented in the previous subsection, the excitation-gap data
$\Delta E$ are readily available.
Remaining task is to perform the numerical derivatives
of 
$\Delta E'$.
We adopted the ``Richardson's deferred approach to 
the limit'' algorithm in the text book \cite{NRF}.
In this algorithm, one takes an extrapolation 
after calculating 
various finite-difference differentiations.
We monitored the relative error, and checked that the error is
kept within $10^{-8}$.
From these preparations, we can calculate 
the $\beta$-function from our simulation data.

\begin{figure}
\includegraphics{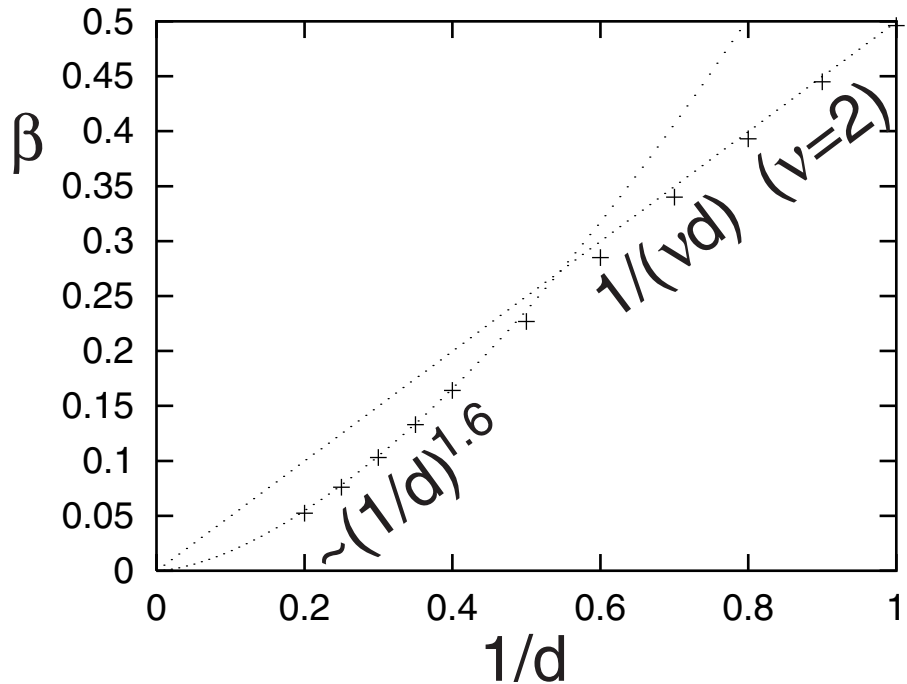}%
\caption{\label{figure_beta_1}
$\beta$-function (\ref{beta_function}) is plotted for $\Sigma=2$.
There appear two regimes with respective asymptotic forms;
see text.
The small-$1/d$ behavior indicates that the transition 
belongs to the universality class of
exponential singularity with $\sigma \approx 0.6$; see Eqs.
 (\ref{beta_exponential}) and (\ref{beta_power}).
}
\end{figure}

We plotted the $\beta$-function $\beta(1/d)$ in Fig. \ref{figure_beta_1}
for $\Sigma=2$.
We see that there appear two regimes:
As is indicated in the plot, in the region $1/d \lesssim 0.3$,
the $\beta$-function is best fitted by a power law $(1/d)^{1.6}$.
Hence, 
the universality would belong to the
exponential singularity (\ref{beta_exponential}) with 
exponent $\sigma=0.6$.
Note that
such stretched-exponential behavior
is suggested in the previous subsection.
On the contrary, the $\beta$-function falls in the simple behavior
$1/\nu (1/d)$ with the index
$\nu=2$ for large $1/d$; see Eq. (\ref{beta_power}).
That result tells that for large $1/d$, the gap formula
enters the regime of power law (\ref{power}).
That is quite convincing, because for large $1/d$,
$\Sigma$ becomes irrelevant and the system reduces \cite{Nishiyama01}
to the
text-book problem of ``particle in a box.''
Hence, the excitation gap is simply
given by the power law (\ref{power})
with the index $\nu=2$.
To summarize, we found that for large $1/d$, the physics changes
so that we must look into the behavior in the vicinity of $1/d=0$.

In order to estimate $\sigma$, we have presented the log-log plot
in Fig. \ref{figure_beta_2} for various $\Sigma$.
From the slopes, we obtain an estimate $\sigma= 0.6(3)$.
This estimate is far less than the mean-field value $\sigma=2$.
It clearly supports
Zaanen's claim $\sigma=2/3$ \cite{Zaanen00,Mukhin01} that
the collision-induced mass gap
is governed by the stretched-exponential singularity.
That is, the collision-induced mass gap is far larger than
that anticipated from meanfield, and hence, 
the string is subjected to robust pinning potential due
to quantum collisions; see Section \ref{section_2}.


Meanwhile,
we had also calculated the Roomany-Wyld approximant of the $\beta$-function
\cite{Roomany80},
\begin{equation}
\label{RW}
\beta= - \frac{1 + \ln(\Delta E_l / \Delta E_{l'})/\ln(l/l')}
 {(\Delta E_l' \Delta E_{l'}'/\Delta E_l\Delta E_{l'})^{1/2}},
\end{equation}
where $l$ and $l'$ denote a pair of system sizes. 
This approximant, in general, exhibits smaller finite-size collections.
In fact,
we found that it converges rapidly to the thermodynamic limit.
However, the resultant data are 
identical to those with the aforementioned formula (\ref{beta})
for sufficiently large $L$.
Because we take advantage of very large system sizes owing to
the density-matrix renormalization group,
the approximant formula (\ref{RW}) is not particularly necessary.

\subsection{Mean-fluctuation deviation}

So far, we have extracted the singularity exponent
directly from the excitation-gap data.
However, as is mentioned in Section \ref{section_2},
the mean deviation of the string undulations
contains an information of the exponent;
\begin{equation}
\label{mean_fluctuation}
\Delta = \sqrt{ \langle x_i^2 \rangle - \langle x_i \rangle^2 } \sim d^{\sigma/2}  .
\end{equation}

\begin{figure}
\includegraphics{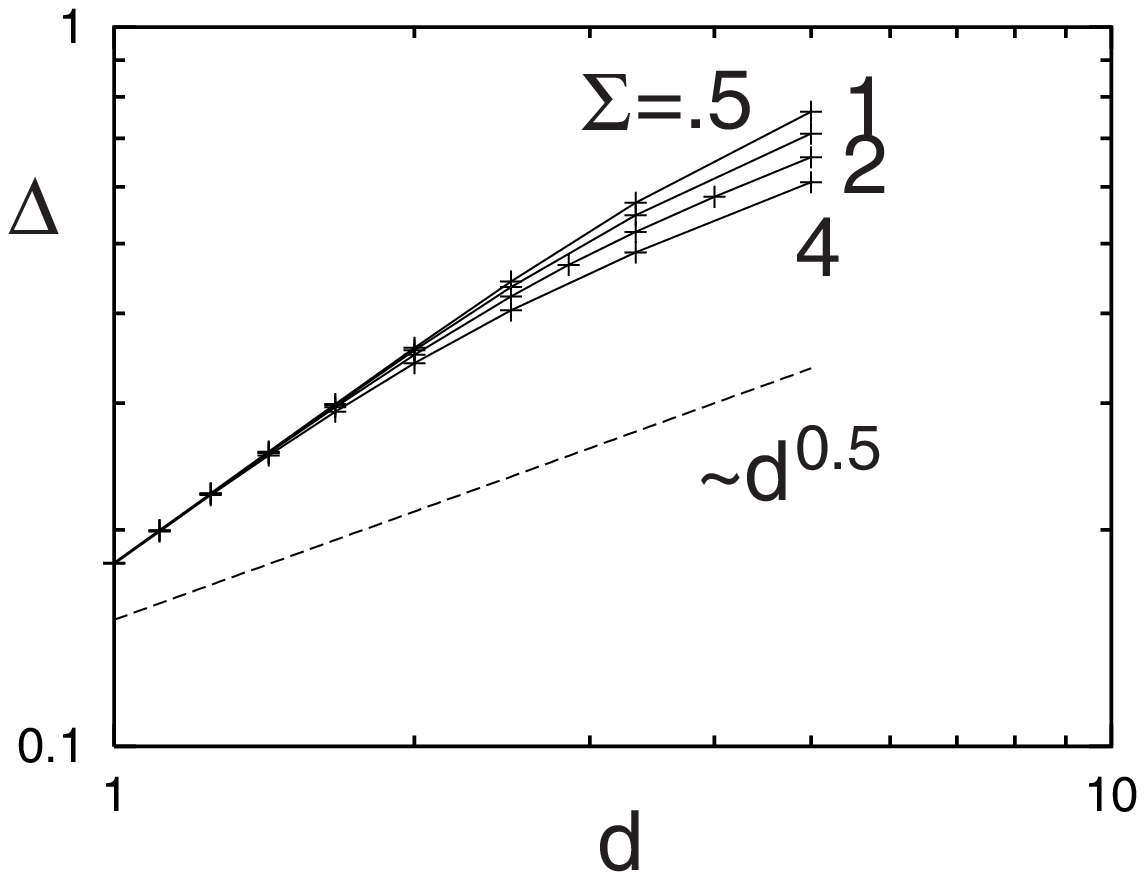}%
\caption{\label{figure_mean_deviation}
Mean fluctuation width
$\Delta$ (\ref{mean_fluctuation})
is plotted for wall spacing $d$.
A naive argument postulates $\Delta \propto d$.
However, the simulation data exhibit
$\Delta \sim d^{0.5}$.
We see that the string is straightened macroscopically.
Our result supports the ``order out of disorder''
mechanism proposed by Refs. \onlinecite{Zaanen00,Mukhin01}.
}
\end{figure}

In Fig. \ref{figure_mean_deviation}, we plotted the mean fluctuation width
$\Delta$ for various $d$ and $\Sigma=0.5$-$4$.
From the plot, we see that there appear two distinctive regimes as in Fig. \ref{figure_beta_1}.
For small $d$, $\Delta$ is proportional to $d$,
as is anticipated from a naive argument; see Section \ref{section_2}.
However, for large $d$, $\Delta$ starts to bend down, indicating that
the string undulation becomes suppressed.
Hence, for large $d$, the conventional picture does not apply.
From the slopes,
we read off the exponent $\sigma/2 \approx 0.5$.
(This result supports our estimate $\sigma=0.6(3)$ 
rather than the meanfield value $\sigma=2$.)
Hence, we see
$\Delta/d \ll 1$ as for $d\to\infty$.
As a consequence, we found
fairly definitely that
the string is straightened macroscopically due to the quantum collisions.
This feature is reminiscent of the infrared divergences
encountered in the theory by Zaanen \cite{Zaanen00}.

\section{\label{section_4}
Summary and discussions}

We have investigated the quantum entropic (collision induced)
interaction 
of an elastic string confined between walls (\ref{Hamiltonian}).
We performed the first-principle simulation
by means of the density-matrix renormalization group.
Our aim is to estimate the singularity exponent $\sigma$
in the excitation-gap formula (\ref{exponential}).
The exponent contains a number of significant informations
which are overviewed in Section \ref{section_2}.

\begin{figure}
\includegraphics{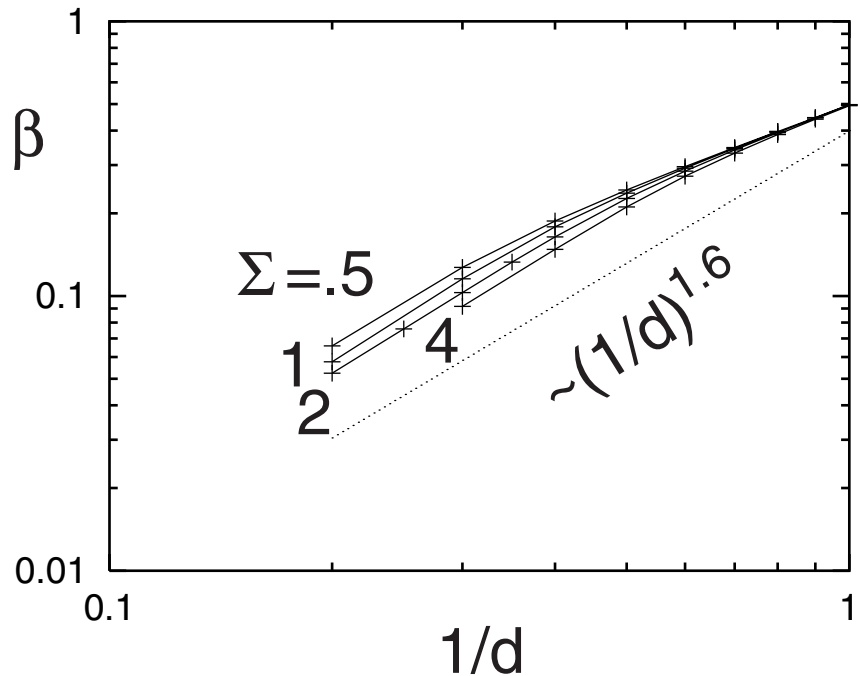}%
\caption{\label{figure_beta_2}
Logarithmic plot of the $\beta$-function for various $\Sigma$.
From the slopes, we read off the singularity exponent $\sigma=0.6(3)$;
see Eq. (\ref{beta_exponential}).
Our result indicates the breakdown of the meanfield picture $\sigma=2$.
}
\end{figure}

In order to estimate $\sigma$, 
we utilized the $\beta$-function (\ref{beta}),
which is readily accessible from our simulation data.
From the asymptotic form of large $d$,
we obtained the estimate
$\sigma = 0.6(3)$ for $\Sigma=0.5$-$4$; 
see Fig. \ref{figure_beta_2}.
Our estimate supports the recent proposal \cite{Zaanen00,Mukhin01}
that $\Delta E$ is described by the stretched exponential form with $\sigma=2/3$,
whereas it might conflict with the meanfield
value $\sigma=2$.
It would be noteworthy that
the elastic modulus also obeys the stretched exponential,
as is shown in our previous study \cite{Nishiyama01}.
Hence,
it is established fairly definitely that
the excitation gap is also described by the stretched exponential
singularity.
Namely,
the excitation gap is much larger than that anticipated from meanfield.
Moreover, we found that
the mean fluctuation width $\Delta$ is far less than
the wall spacing $d$; namely $\Delta \sim d^{0.5}$.
This again supports our conclusion rather than the meanfield;
see Eq. 
(\ref{deviation_and_exponent}).

From those observations, we are led to the conclusion that
the string is confined by extremely robust effective pinning potential due
to the quantum collisions; subsequently, large mass gap opens and $\Delta$ gets bounded.
Those features are also quite consistent with the findings of the series of works
\cite{Zaanen00,Mukhin01}, validating their
``order out of disorder'' mechanism
responsible to the stabilization of the stripe phase.


Provided that plural strings are concerned,
does the value $\sigma$ get affected?
As a matter of fact,
the exponent $\sigma$ is not necessarily universal, but
it does exhibit continuous variation
\cite{Young79,Nelson79} for a certain class of extended models.
More specifically, in Refs. \onlinecite{Itoi01,Itoi99}, the authors claim
that for extended models
with high central charges (that is, many random surfaces) 
 exhibit various types of criticalities.
Therefore,
It is likely that $\sigma$ would acquire corrections for multi-string case.
Extended simulation directed to this issue would be
remained for future study.


%

%



\begin{acknowledgments}
This work is supported by Grant-in-Aid for
Scientific Research Program
(No. 13740240) from Monbusho, Japan.
The author is grateful to Dr. T. Momoi
for helpful discussions.
\end{acknowledgments}


\end{document}